\def\lb{\hfil\break}
\def\pmb#1{\setbox0=\hbox{#1}%
   \kern-.025em\copy0\kern-\wd0
   \kern.05em\copy0\kern-\wd0
   \kern-0.025em\raise.0433em\box0}
\def\gta{\mathrel{{\lower 3pt\hbox{$\mathchar"218$}}\hskip-8pt
   \raise 2pt\hbox{$\mathchar"13E$}}}
\def\lta{\mathrel{{\lower 3pt\hbox{$\mathchar"218$}}\hskip-8pt
   \raise 2pt\hbox{$\mathchar"13C$}}}
\def\half{{\scriptstyle{1\over2}}}
\def\dagg{\phantom{\dagger}}            
\def\subboldc{\pmb{$\scriptstyle c$}}   
\def\boldd{\pmb{d}}                     
\def\boldell{\pmb{$\ell$}}              
\def\subboldell{\pmb{$\scriptstyle\ell$}}
\def\boldT{\pmb{$T$}}                   
\def\bolddelta{\pmb{$\delta$}}
\def\subbolddelta{\pmb{$\scriptstyle\delta$}}
\def\boldeta{\pmb{$\eta$}}
\def\subboldeta{\pmb{$\scriptstyle\eta$}}
\def\boldxi{\pmb{$\xi$}}
\def\vacuum{|\pmb{\it O}\thinspace\rangle}
\def\up{\uparrow}
\def\dn{\downarrow}
\def\ud{\uparrow\downarrow}
\def\today{\number\day\space\ifcase\month\or
  January\or February\or March\or April\or May\or June\or
  July\or August\or September\or October\or November\or December\fi
 \space\number\year}
\documentstyle[twocolumn,aps]{revtex}
\font\tenrm=cmr8
\begin{document}
\newcommand{\Vv }{{\raisebox{-1.2pt}{\makebox(0,0){$o$}}}}
\newcommand{\Zz }{{\raisebox{-1.2pt}{\makebox(0,0){$\mbox{\tiny o}$}}}}
\newcommand{\Xx }{{\special{em:moveto}}}
\newcommand{\Yy }{{\special{em:lineto}}}
\newcommand{\Ww }{{\usebox{\plotpoint}}}
\title{\begin{minipage}{480pt}
 {\huge \ \\ \ \\ \ \\  \ \\
Stripes, Non-Fermi-Liquid Behavior, and High-$T_c$ \\
Superconductivity \\ 
\ \\}
{\large \bf J. Ashkenazi$^1$\ \\ }
\end{minipage}
}
\author{\ \hfill
\begin{minipage}{5in}
\small
\baselineskip 11pt
\underline{\it Received 19 February 1997 \ \ \ \ \ \ \ \ \ \ \ \ \ \
\ \ \ \ \ \ \ \ \ \ \ \ \ \ \ \ \ \ \ \ \ \ \ \ \ \ \ \ \ \ \ \ \ \ \ \
\ \ \ \ \ \ \ \ \ \ \ \ \ \ \ \ \ \ \ \ \ \ \ \ \ } \\
The electronic structure of the high-$T_c$ cuprates is studied in terms
of ``large-$U$'' and ``small-$U$'' orbitals. A striped structure and
three types of quasiparticles are obtained, polaron-like ``stripons''
carrying charge, ``svivons'' carrying spin, and ``quasielectrons''
carrying both. The anomalous properties are explained, and specifically 
the behavior of the resistivity, Hall constant, and thermoelectric
power. High-temperature super- 
\underline{conductivity results from transitions between pair
states of quasielectrons and stripons. } 
\end{minipage}
}
\maketitle
\setlength{\unitlength}{1in}
\footnotetext[1]{\footnotesize Physics Department, University of Miami,
P.O. Box 248046, Coral Gables, FL 33124, U.S.A.} 
\makeatletter
\global\@specialpagefalse
\def\@oddhead{\footnotesize \it Journal of Superconductivity,
Vol. 10, No. ?, 1997 \hfill \ }
\makeatother
\baselineskip 12pt
\normalsize \rm
\par\noindent{\bf 1. INTRODUCTION}
\medskip

The validity of the Fermi-liquid scenario for the high-$T_c$ cuprates
has been doubted, and scenarios such as of a marginal Fermi liquid (MFL)
[1], anomalously-screened Fermi liquid [2], and others have been
suggested. Evidence is growing [3,4] that the CuO$_2$ planes are
characterized by a static or dynamic striped structure. A theoretical
study of the cuprates requires the consideration of both ``large-$U$''
and ``small-$U$'' orbitals [5]. 

A small-$U$ electron in band $\nu$, spin $\sigma$, and wave vector ${\bf
k}$ is created by the fermion operators $c_{\nu\sigma}^{\dagger}({\bf
k})$. The large-$U$ states are treated by the ``slave-fermion'' method
[6], where an electron in site $i$ and spin $\sigma$ is created by
$d_{i\sigma}^{\dagger} = e_i^{\dagger} s_{i,-\sigma}^{\dagg}$, if it is
in the ``upper-Hubbard-band'', and by $d_{i\sigma}^{\prime\dagger} =
\sigma s_{i\sigma}^{\dagger} h_i^{\dagg}$, if it is in a Zhang-Rice-type
``lower-Hubbard-band''. Here $e_i^{\dagg}$ and $h_i^{\dagg}$ are
(``excession'' and ``holon'') fermion operators, and
$s_{i\sigma}^{\dagg}$ are (``spinon'') boson operators. The constraint
$e_i^{\dagger} e_i^{\dagg} + h_i^{\dagger} h_i^{\dagg} + \sum_{\sigma}
s_{i\sigma}^{\dagger} s_{i\sigma}^{\dagg} = 1$ should be satisfied in
every site. 

An auxiliary Hilbert space is introduced where the constraint is imposed
only on the average by introducing a chemical-potential-like Lagrange
multiplier. Physical observables are projected into the physical Hilbert
space by taking appropriate combinations of Green's functions of the
auxiliary space. Since the time evolution of Green's functions is
determined by the Hamiltonian which obeys the constraint rigorously,
effects of constraint violation may result only from approximations
introduced. Within the ``spin-charge separation'' approximation
two-particle spinon-holon Green's functions are decoupled. 

\bigskip
\par\noindent{\bf 2. QUASIPARTICLES}
\medskip

The spinons are diagonalized by the Bogoliubov transformation, yielding
creation operators $\zeta_{\sigma}^{\dagger} ({\bf k})$, and ``bare''
spinon energies $\epsilon^{\zeta} ({\bf k})$ with a V-shape zero minimum
at ${\bf k}={\bf k}_0$, where ${\bf k}_0$ is either $( {\pi \over
2{\rm a}} , {\pi \over 2{\rm a}} )$ or $( {\pi \over 2{\rm a}} , -{\pi
\over 2{\rm a}} )$. Bose condensation results in antiferromagnetism
(AF), and the spinon reciprocal lattice is extended by adding the wave
vector ${\bf Q}=2{\bf k}_0$. 

The decoupling of two-particle spinon-spinon Green's functions, relevant
for spin processes, is more reasonable within the slave-fermion method,
where the Bose condensation of spinons does not require pair
correlation, than within the ``slave-boson'' [6] method, where BCS
condensation of spinons does require pair correlation. 

A lightly doped AF plane tends to separate into a ``charged'' phase and
an AF phase. Under long-range Coulomb interactions one expects [7] a
frustrated striped structure of these phases. Experiment [4] confirms
such a scenario and indicates at least in certain cases a structure
where narrow charged stripes form antiphase domain walls separating 
wider AF stripes. Various experiments [8] support the assumption that
such a structure exists, at least dynamically, in all the
superconducting cuprates. 

The validity of the spin-charge separation approximation has been
established in one-dimension. Thus it should apply for holons
(excessions) within the charged stripes, and they are referred to as
``stripons''. Their fermion creation operators are denoted by
$p^{\dagger}_{\mu}({\bf k})$, and their bare energies by
$\epsilon^p_{\mu}({\bf k})$. Since one expects finite stripe segments,
frustrations, and defects, which are fatal for itinerancy in
one-dimension, it is likely that the starting point for the stripon
states is of localized states.

The small-$U$ electrons hybridize with coupled holon-spinons
(excession-spinons) within the AF stripes forming, within the auxiliary
space,``Quasi-electrons'' (QE's), created by
$q_{\iota\sigma}^{\dagger}({\bf k})$. Their bare energies
$\epsilon^q_{\iota} ({\bf k})$ form quasi-continuous ranges of bands
crossing the Fermi level ($E_{_{\rm F}}$) over ranges of the Brillouin
zone (BZ). 

\makeatletter
\global\@specialpagefalse
\def\@oddhead{\ \\ \ \hfill \bf Ashkenazi}
\makeatother

\bigskip
\par\noindent{\bf 3. SPECTRAL FUNCTIONS}
\medskip

The electron spectral function $A({\bf p}, \omega)$ is expressed in
terms of spectral functions $A^q_{\iota}({\bf k}, \omega)$,
$A^{\zeta}_{\lambda}({\bf k}, \omega)$, and $A^p_{\mu}({\bf k},
\omega)$, of the QE's, spinons, and stripons, respectively. $A({\bf p},
\omega)$ has a ``coherent'' contribution from a few QE bands, while the
quasi-continuum of the other QE bands and the stripon-spinon spectral
functions contribute an ``incoherent'' background of a comparable 
integrated weight.

The quasiparticles are coupled by a Hamiltonian term, derived from
hopping and hybridization terms of the original Hamiltonian, and
expressed as: 
\begin{eqnarray}
{\cal H}^{\prime} &=& {1 \over \sqrt{N}} \sum_{\iota\mu\lambda\sigma}
\sum_{{\bf k}, {\bf k}^{\prime}} \Big\{\sigma
\epsilon^{qp}_{\iota\mu\lambda\sigma}({\bf k}^{\prime}, {\bf k})
q_{\iota\sigma}^{\dagger}({\bf k}) p_{\mu}^{\dagg}({\bf k}^{\prime})
\nonumber \\ &\ &\times\big[ \cosh{(\xi_{\lambda\sigma,({\bf k} - {\bf
k}^{\prime})})} \zeta_{\lambda\sigma}^{\dagg}({\bf k} - {\bf
k}^{\prime}) \nonumber \\ &\ &+ \sinh{(\xi_{\lambda\sigma,({\bf k} -
{\bf k}^{\prime})})} \zeta_{\lambda,-\sigma}^{\dagger}({\bf k}^{\prime}
- {\bf k}) \big] + h.c. \Big\}, 
\end{eqnarray} 
where the $\cosh{}$ and $\sinh{}$ terms are those appearing in the
Bogoliubov transformation. ${\cal H}^{\prime}$ introduces a vertex
connecting QE, stripon and spinon propagators. Since the stripon
bandwidth turns out to be much smaller than the QE and spinon
bandwidths, ``vertex corrections'' are negligible by a generalized
Migdal theorem, and a second-order perturbation expansion in ${\cal
H}^{\prime}$ is applicable. The scattering rates $\Gamma^q_{\iota}({\bf
k}, \omega)$, $\Gamma^{\zeta}_{\lambda}({\bf k}, \omega)$, and
$\Gamma^p_{\mu}({\bf k}, \omega)$, of the quasiparticles are then
calculated, and for sufficiently doped cuprates one gets a
self-consistent solution of the following features: 

\underbar{Spinons}: One gets $A^{\zeta}({\bf k},
\omega)\propto\omega$ for small $\omega$, and thus $A^{\zeta}({\bf
k}, \omega) b_{_T}(\omega)\propto T$ for $\omega\ll T$, where
$b_{_T}(\omega)$ is the Bose distribution function. 

\underbar{Stripons}: The localized stripon states are renormalized to
polaron-like states very close to $E_{_{\rm F}}$, with some hopping
through QE-spinon states. One gets $\Gamma^p({\bf k}, \omega) \propto A
\omega^2 + B \omega T + CT^2$, and a two-dimensional itinerant behavior
at low temperatures, with a bandwidth of $\sim$$0.02\;$eV. 

\underbar{Quasi-electrons}: An approximate expression for their
scattering rates is given by $\Gamma^q({\bf k}, \omega) \propto
\omega[b_{_T}(\omega) + \half]$, becoming $\Gamma^q({\bf k},
\omega)\propto T$ in the limit $T\gg |\omega|$, and 
$\Gamma^q({\bf k}, \omega)\propto\half |\omega|$ in the limit
$T\ll |\omega|$, in agreement with MFL phenomenology [1]. 

\underbar{Lattice effects (``svivons'')}: The charged stripes are
characterized by an LTT-like structure [3]. Thus, spinon excitations due
to ${\cal H}^{\prime}$ are followed by phonon excitations, and stripons
have polaron-like lattice features. A spinon propagator linked to a
vertex is thus ``dressed'' by phonon propagators. We refer to such a
phonon-dressed spinon as a svivon. 

\underbar{Optical conductivity}: QE $\to$ QE transitions result in the
observed Drude peak [9], while stripon $\to$ stripon-svivons and stripon
$\to$ QE-svivon transitions result in the observed mid-IR peak [9]. 

\underbar{Spectroscopic anomalies}: ``Shadow bands'', ``extended'' van
Hove singularities (vHs), and normal-state pseudogaps result from the
effect of the striped structure on the QE bands [10]. The vHs are
extended to supply the spectral weight for the stripon states, and when
the vHs are missing this transferred spectral weight causes a pseudogap
in the same place in the BZ [11]. The spectroscopic signature of
stripons is smeared over few tenths of an eV around $E_{_{\rm F}}$ due
to the accompanying svivon excitations. 

\bigskip
\par\noindent{\bf 4. TRANSPORT PROPERTIES}
\medskip

The dc current is expressed as a sum ${\bf j} = {\bf j}^q + {\bf j}^p$
of QE and stripon contributions. Since stripons do not hop directly, but
via QE states, one gets that ${\bf j}^p \cong \alpha {\bf j}^q$, where
$\alpha$ is approximately $T$-independent. Consequently, an electric
field is accompanied by gradients {\bf \pmb{$\nabla$}\/}$\mu^q$ and {\bf
\pmb{$\nabla$}\/}$\mu^p$ of the QE and stripon chemical potentials, such
that $N^q${\bf \pmb{$\nabla$}\/}$\mu^q + N^p${\bf
\pmb{$\nabla$}\/}$\mu^p = 0$, where $N^q$ and $N^p$ are the
contributions of QE's and stripons to the electron density of states at
$E_{_{\rm F}}$. 

Expressions for the dc conductivity and Hall constant, in terms of
Green's functions, are derived using the Kubo formalism. They are
expressed through diagonal and non-diagonal conductivity QE and stripon
terms $\sigma_{xx}^{qq}$, $\sigma_{xx}^{pp}$, $\sigma_{xy}^{qqq}$,
$\sigma_{xy}^{ppp}$, and mixed terms $\sigma_{xy}^{qqpp}$. The currents
in an electric field ${\bf E}$ can then be expressed as:
$j_x^q=\sigma_{xx}^{qq} {\cal E}_x^q$, $j_x^p=\sigma_{xx}^{pp}
{\cal E}_x^p$, where {\bf \pmb{${\cal E}$}\/}$^q={\bf E} + ${\bf
\pmb{$\nabla$}\/}$\mu^q/{\rm e}$, {\bf \pmb{${\cal E}$}\/}$^p={\bf
E} + ${\bf \pmb{$\nabla$}\/}$\mu^p/{\rm e}$. By expressing ${\bf
E}=(N^q ${\bf \pmb{${\cal E}$}\/}$^q + N^p $ {\bf \pmb{${\cal
E}$}\/}$^p)/(N^q+N^p)$, and $j^q_x + j^p_x = j_x = E_x/\rho_x$, one gets
that the resistivity can be expressed as: 
\begin{equation}
\rho_x = {1 \over (N^q+N^p) (1+\alpha)} \Big( {N^q \over
\sigma_{xx}^{qq}} + {\alpha N^p \over \sigma_{xx}^{pp}}\Big).  
\end{equation}
Similarly, the Hall constant $R_{_{\rm H}} = E_y/j_xH$ can be 
expressed as $R_{_{\rm H}} = \rho_x / \cot{\theta_{_{\rm H}}}$, 
where:
\begin{equation}
{\cot{\theta_{_{\rm H}}} \over (1+\alpha)} = \Big[{\sigma_{xy}^{qqq} +
\sigma_{xy}^{qqpp} \over \sigma_{xx}^{qq}} + {\alpha(\sigma_{xy}^{ppp} +
\sigma_{xy}^{qqpp}) \over \sigma_{xx}^{pp}} \Big]^{-1}. 
\end{equation}

\vskip -1.1truecm
\begin{figure}[t]
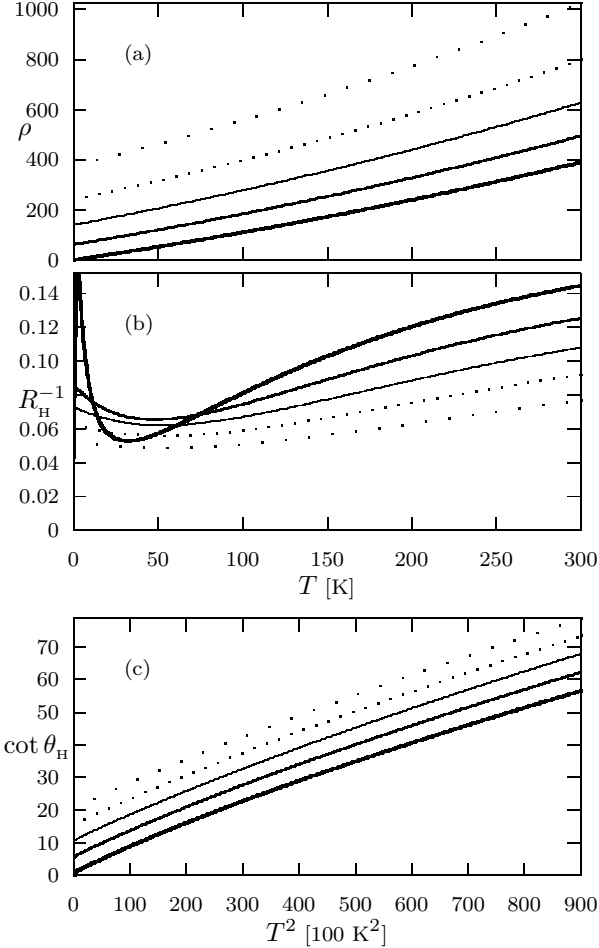


\setlength{\unitlength}{0.240900pt}
\ifx\plotpoint\undefined\newsavebox{\plotpoint}\fi
\sbox{\plotpoint}{\rule[-0.175pt]{0.350pt}{0.350pt}}%


\caption{The resistivity (a), inverse Hall constant (b), and
$\cot{\theta_{_{\rm H}}}$ (c), in arbitrary unit, for parameter values:
A=1,7,13,19,25; B=.001; C=1; D=0,50,100,150,200; N=1,.9,.8,.7,.6; Z=2.
The first value corresponds to the thickest lines.} 
\label{F1}
\end{figure}

The temperature dependencies are determined by these of $\Gamma^q$ and
$\Gamma^p$, obtained above, to which we add temperature-independent
impurity scattering terms. Thus one can parametrize:
$\sigma_{xx}^{qq}\propto (D+CT)^{-1}$, $\sigma_{xx}^{pp}\propto
(A+BT^2)^{-1}$, $\sigma_{xy}^{qqq}\propto (D+CT)^{-2}$,
$\sigma_{xy}^{ppp}\propto (A+BT^2)^{-2}$, $\sigma_{xy}^{qqpp}\propto
[(D+CT)(A+BT^2)]^{-1}$, and express: 
\begin{eqnarray}
\rho_x &\cong& {(D+CT+A+BT^2) \over N}, \\ \cot{\theta_{_{\rm H}}}
&\cong& \Big({Z \over D+CT} + {1 \over A+BT^2} \Big)^{-1}. 
\end{eqnarray} 

\vskip -1.1truecm
\begin{figure}[t]
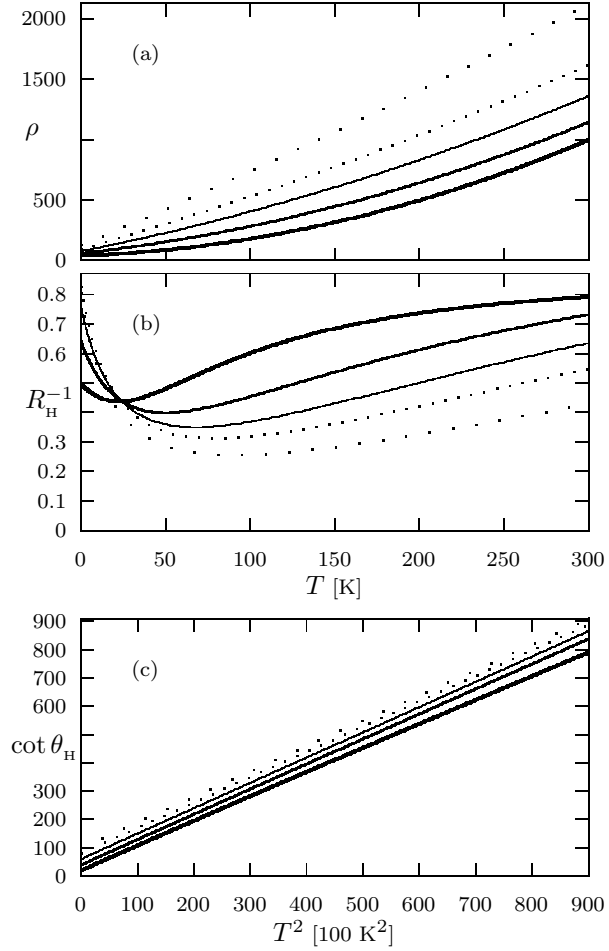


\setlength{\unitlength}{0.240900pt}
\ifx\plotpoint\undefined\newsavebox{\plotpoint}\fi
\sbox{\plotpoint}{\rule[-0.175pt]{0.350pt}{0.350pt}}%


\caption{The resistivity (a), inverse Hall constant (b), and
$\cot{\theta_{_{\rm H}}}$ (c), in arbitrary unit, for parameter values:
A=20,40,60,80,100; B=.01; C=.5,2,5,10,20; D=20,40,80,160,320;
N=1,1.3,1.8,2.5,3.4; Z=.01. The first value corresponds to the thickest
lines.} 
\label{F2}
\end{figure}

\makeatletter
\global\@specialpagefalse
\def\@oddhead{Stripes, non-Fermi-liquid behavior \hfill \ }
\makeatother

This parametrization reproduces the systematic behavior of the transport
quantities in different cuprates, except for the effect of the pseudogap
for underdoped cuprates. Results for sets of parameters corresponding to
data in YBa$_2$Cu$_{3-x}$Zn$_x$O$_7$ [12], Tl$_2$Ba$_2$CuO$_{6+\delta}$
[13], and La$_{2-x}$Sr$_x$CuO$_4$ [14], are presented in Figs. 1, 2, and
3, respectively. 
 
When also a temperature gradient is present, one can express: ${\bf
j}^q={\rm e}T^{-1} \underline{\bf L}^{q(11)} ${\bf \pmb{${\cal
E}$}\/}$^q +\underline{\bf L}^{q(12)} ${\bf \pmb{$\nabla$}\/}$(T^{-1})$,
${\bf j}^p={\rm e}T^{-1} \underline{\bf L}^{p(11)} ${\bf \pmb{${\cal
E}$}\/}$^p +\underline{\bf L}^{p(12)} ${\bf \pmb{$\nabla$}\/}$(
T^{-1})$. The thermoelectric power (TEP) is given by $\underline{\bf
S}=[{\bf E} / ${\bf \pmb{$\nabla$}\/}$T]_{{\bf j} = 0}$. Since ${\bf
j}^p\cong\alpha {\bf j}^q$, the condition ${\bf j}=0$ means
${\bf j}^q\cong{\bf j}^p\cong 0$. Thus, one can express:
$\underline{\bf S}=(N^q\underline{\bf S}^q + N^p\underline{\bf S}^p)
/ (N^q + N^p)$, where $\underline{\bf S}^q=- \underline{\bf
L}^{q(12)} / {\rm e}T \underline{\bf L}^{q(11)}$, $\underline{\bf
S}^p=- \underline{\bf L}^{p(12)} / {\rm e}T \underline{\bf
L}^{p(11)}$. 

\vskip -0.7truecm
\begin{figure}[t]
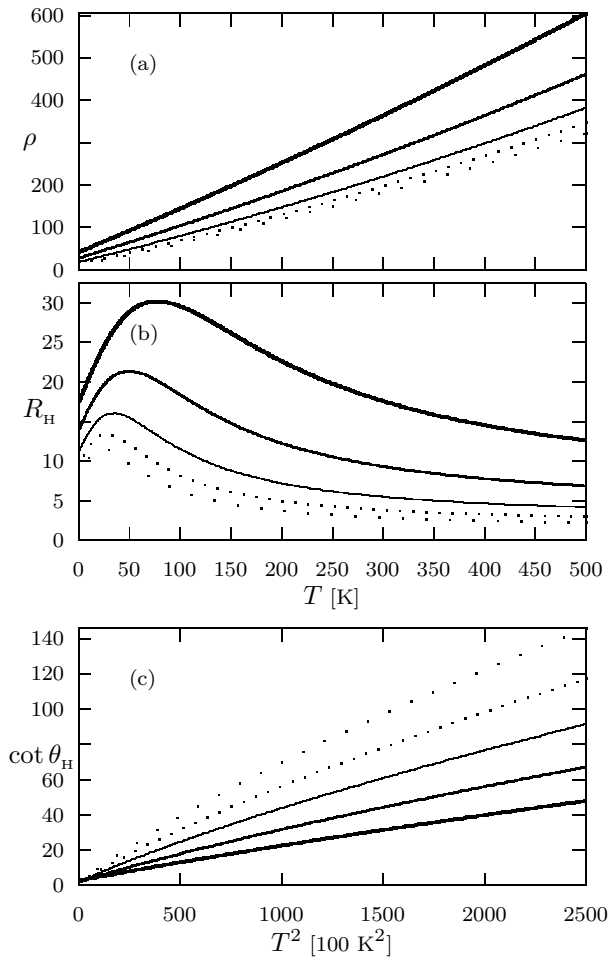


\setlength{\unitlength}{0.240900pt}
\ifx\plotpoint\undefined\newsavebox{\plotpoint}\fi
\sbox{\plotpoint}{\rule[-0.175pt]{0.350pt}{0.350pt}}%


\caption{The resistivity (a), Hall constant (b), and $\cot{\theta_{_{\rm
H}}}$ (c), in arbitrary unit, for parameter values: A=3,2.4,2,1.7,1.5;
B=.00025,.0004,.0006,.0008,.001; C=1,1.1,1.3,1.6,2; D=40;
N=1,1.5,2.2,3,4; Z=3. The first value corresponds to the thickest
lines.} 
\label{F3}
\end{figure}

One gets $S^q\propto T$, as for wide-band electrons, while $S^p$
saturates at $T \simeq 200\; $K to the narrow-band result:
$S^p=(k_{_{\rm B}}/{\rm e})\ln{[(1}-n^p)/n^p]$, where $n^p$ is the
fractional occupation of the stripon band. This is consistent with the
typical behavior of the TEP in the cuprates which has been parametrized
as: $S=AT+BT^{\alpha} / (T+\Theta)^{\alpha}$ [15]. It was found [16,17]
that $S^p=0$ (and thus $n^p=0.5$) close to the optimal stoichiometry. 

\bigskip
\par\noindent{\bf 5. MECHANISM FOR HIGH-\pmb{$T_c$}\/}
\medskip

${\cal H}^{\prime}$ provides a mechanism for high-$T_c$ due to
transitions between pair states of QE's and stripons through the
exchange of svivons. The symmetry of the superconducting gap is
determined by the symmetry of coupling through ${\cal H}^{\prime}$, and
is thus close to the symmetry of the normal-state pseudogap, as has been
observed [11]. This pairing mechanism is similar to the interband pair
transition mechanism proposed by Kondo [18]. An upper limit for $T_c$ is
determined by the temperature where the stripon band becomes coherent,
and similarly to a previous work [5], this turns out to be consistent
with the Uemura limit. 
 
\bigskip
\par\noindent{\bf 6. SUMMARY}
\medskip

The consideration of large-$U$ and small-$U$ orbitals in the cuprates
results in a striped structure, and three types of quasiparticles:
polaron-like stripons carrying charge, phonon-dressed spinons (svivons)
carrying spin, and QE's carrying both. Anomalous normal-state properties
of the cuprates are explained, and specifically the systematic behavior
of the resistivity, Hall constant, and thermoelectric power. A mechanism
for high-$T_c$ is obtained on the basis of transitions between pair
states of stripons and QE's through the exchange of svivons. 

\makeatletter
\global\@specialpagefalse
\def\@oddhead{\ \\ \ \hfill \bf Ashkenazi}
\makeatother


\makeatletter
\global\@specialpagefalse
\makeatother


\bigskip
{\bf REFERENCES \hfill } \\

\vskip -0.15truecm
\baselineskip 10pt
\footnotesize
\noindent
1. \ C.~M.~Varma {\it et al.}, {\it Phys.~Rev.~Lett.} {\bf 63}, 1996
(1989). \\ 
2. \ M.~Weger, {\it et al.}, {\it Z.~Phys.~B} {\bf 101}, 573 (1996). \\
3. \ A.~Bianconi {\it et al.}, {\it Phys.~Rev.~B} {\bf 54}, 12018, (1996); 
\\ \mbox{\ } \ \ \ {\it Phys.~Rev.~Lett.} {\bf 76} 3412 (1996). \\ 
4. \ J.~M.~Tranquada {\it et al.}, {\it Phys.~Rev.~B} {\bf 54}, 7489, (1996);
\\ \mbox{\ } \ \ \ {\it Phys.~Rev.~Lett.} {\bf 78}, 338 (1997). \\ 
5. \ J.~Ashkenazi, {\it J.~Supercond.} {\bf 7}, 719 (1994); {\it ibid}
{\bf 8}, 559 \\ \mbox{\ } \ \ \ (1995). \\ 
6. \ S.~E.~Barnes, {\it Adv.~Phys.} {\bf 30}, 801 (1980). \\
7. \ V.~J.~Emery, and S.~A.~Kivelson, {\it Physica C} {\bf 209}, 597 
\\ \mbox{\ } \ \ \ (1993). \\
8. \ Papers in this issue. \\
9. \ D.~B.~Tanner, and T.~Timusk, {\it Physical Properties \\ \mbox{\ }
\ \ \ of High Temperature Superconductors III}, edited by \\ \mbox{\ } \
\ \ D.~M.~Ginsberg (World Scientific, 1992), p. 363. \\ 
10. M.~I.~Salkola, {\it et al.}, {\it Phys.~Rev.~Lett.} {\bf 77}, 155
(1996). \\ 
11. D.~S.~Marshall, {\it et al.}, {\it Phys.~Rev.~Lett.} {\bf 76}, 4841
(1996). \\ 
12. T.~R.~Chien, {\it et al.}, {\it Phys.~Rev.~Lett.} {\bf 67}, 2088
(1991). \\ 
13. Y.~Kubo and T.~Manako, {\it Physica C} {\bf 197}, 378 (1992). \\
14. H.~Takagi, {\it et al.}, {\it Phys.~Rev.~Lett.} {\bf 69}, 2975
(1992); \\ \mbox{\ } \ \ \ H.~Y.~Hwang, {\it et al.}, {\it ibid.} {\bf
72}, 2636 (1994). \\ 
15. S. Tanaka, {\it et al.}, {\it J.~Phys.~Soc.~Japan} {\bf 61}, 1271 
(1992). \\
16. B.~Fisher, {\it et al.}, {\it J. Supercond.} {\bf 1}, 53 (1988);
J. Genossar, \\ \mbox{\ } \ \ \ {\it et al.}, {\it Physica C} {\bf
157}, 320 (1989). \\ 
17. K.~Matsuura, {\it et al.}, {\it Phys.~Rev.~B} {\bf 46}, 11923
(1992); \\ \mbox{\ } \ \ \ \ S.~D.~Obertelli, {\it et al.}, {\it ibid.},
p. 14928; C.~K.~Subramaniam, \\ \mbox{\ } \ \ \ {\it et al.}, {\it
Physica C} {\bf 203}, 298 (1992). \\ 
18. J.~Kondo, {\it Prog.~Theor.~Phys.} {\bf 29}, 1 (1963). \\ 
 
\end{document}